\documentclass[12pt]{article}
\usepackage{cite}
\usepackage{color}
\usepackage{graphicx}

\makeatletter
\@addtoreset{equation}{section}

\makeatletter
\renewcommand\section{\@startsection {section}{1}{\z@}%
                                   {-3.5ex \@plus -1ex \@minus -.2ex}
                                   {2.3ex \@plus.2ex}%
                                   {\normalfont\large\bfseries}}
\renewcommand\subsection{\@startsection{subsection}{2}{\z@}%
                                     {-3.25ex\@plus -1ex \@minus -.2ex}%
                                     {1.5ex \@plus .2ex}%
                                     {\normalfont\bfseries}}

\def\baselinestretch{1.2}
\newcommand{\be}{\begin{equation}}
\newcommand{\ee}{\end{equation}}
\newcommand{\beq}{\begin{eqnarray}}
\newcommand{\eeq}{\end{eqnarray}}

\begin{document}
\begin{titlepage}
\begin{flushright}
hep-th/0410123\\
MAD-TH-04-11
\end{flushright}

\vfil\

\begin{center}

{\Large{\bf Dualities, Twists, and Gauge Theories\\ with Non-Constant Non-Commutativity}}

\vfil

Akikazu Hashimoto and Keith Thomas

\vfil

Department of Physics\\
University of Wisconsin\\ Madison, WI 53706\\

\vfil

\end{center}

\begin{abstract}
\noindent We study the world volume theory of D3-branes wrapping the
Melvin universe supported by background NSNS $B$-field.  In the
appropriate decoupling limit, the open string dynamics is that of
non-commutative guage field theory with non-constant
non-commutativity. We identify this model as a simple Melvin twist of
flat D3 branes. Along similar lines, one recognizes the model of
Hashimoto and Sethi as being the Melvin null twist, and the model of
Dolan and Nappi as being the null Melvin twist, of the flat
D3-brane. This construction therefore offers a unified perspective on
most of the known explicit constructions of non-commutative gauge
theories as a decoupled theory of D-branes in a $B$-field background.
We also describe the world volume theory on the D3-brane in Melvin
universe which is decaying via the nucleation of monopole
anti-monopole pair.
\end{abstract}
\vspace{0.5in}

\end{titlepage}
\renewcommand{\baselinestretch}{1.05}  

\section{Introduction}

Non-commutative gauge theory is an intriguing dynamical system which
exhibits rich features such as gauge invariance, non-locality and UV/IR
mixing. These are features commonly seen in more sophisticated theories
such as string theory, little string theory, and gravity.  Yet, they
are simple enough to admit a Lagrangian formulation and for many
purposes can be treated as an ordinary field theory.  For this reason,
they serve as useful toy model.

Non-commutative gauge theories have become the focus of intense
investigation in the recent years in light of the realization that
they arise naturally as a decoupling limit of open string dynamics in
the presence of a background $B$-field
\cite{Connes:1997cr,Douglas:1998fm,Seiberg:1999vs}. So far, most of
the work in this direction has focused on non-commutative gauge
theories with constant non-commutativity parameter $\theta^{\mu \nu}$.
Generalization of the Moyal $*$-product to general Poisson manifolds
\cite{Kontsevich:1997vb} and its relation to certain topological sigma
model \cite{Cattaneo:1999fm} are well known. Very interesting early
work relating open string dynamics on weakly varying $B$ field to
non-constant non-commutativity parameter can be found in
\cite{Cornalba:2001sm}. Nontheless, only a handful of concrete string
theory realizations of gauge theories with non-constant
non-commutativity are known
\cite{Hashimoto:2002nr,Dolan:2002px,Lowe:2003qy}.

In this article, we show that a large class of non-commutative gauge
theories with non-constant non-commutativity can be constructed by
applying a sequence of duality transformations and twists to flat
D-branes.  The prototype of this construction is a D-brane wrapping a
Melvin universe supported by the flux of an NSNS B-field. We refer to
such sequences of dualities and twists as {\it Melvin twists}.  The
sequence of steps is very similar to the ones used to construct dipole
theories
\cite{Bergman:2000cw,Bergman:2001rw,Ganor:2002ju,Alishahiha:2003ru}
but the orientation of the branes is different.  Similar methods were
used to construct supergravity solutions of Taub-NUT geometry in a
non-trivial $B$-field background \cite{Dasgupta:2003us}, as well as
black branes \cite{Gimon:2003xk} and Aichelburg-Sexel waves \cite{Nastase} in
various asymptotic geometries. Slight variation in the construction of
decoupled gauge theories can be characterized by the variation in the
Melvin twists. In fact, we find that most of the known examples of
non-commutative gauge theories with non-constant non-commutativity can
be realized as a Melvin twist of a flat D-brane.

One interesting property of Melvin backgrounds in string theory is the
fact that they are generically non-supersymmetric and can decay via
nucleation of monopole anti-monopole pairs
\cite{Dowker:1994bt,Dowker:1994up,Dowker:1995gb,Dowker:1996sg}.  In
light of the fact that the world volume theory of D-branes in a Melvin
background is a non-commutative field theory, it is natural to expect
that the world volume theory of D-branes in a decaying Melvin
background would be some sort of non-commutative gauge theory with
explicit time dependence. This scenario will also be explored in this
paper.

The organization of this paper is as follows. We will begin in section
2 by formulating the decoupling limit of D-branes in a Melvin universe
with background NSNS 3-form flux.  This example will serve as the
prototype for the Melvin twist construction of non-commutative gauge
theories.  In section 3, we explain how the other explicitly known
examples of non-commutative gauge theories, including the models of
\cite{Hashimoto:2002nr} and \cite{Dolan:2002px} can be viewed as a
variant of the Melvin twist of flat D-branes. In section 4, we
describe how the decay of Melvin space-time via nucleation of monopole
anti-monopole pairs affects the world volume theory on the brane. We
present our conclusions in section 5.

\section{D-branes in  Melvin universe}

In this section we will describe the effective world volume dynamics
of D-branes in a Melvin universe supported by NSNS 3-form field
strength. As this example will serve as a prototype of the
duality/twist construction of non-commutative gauge theories, we will
describe the steps of the construction in some detail.

Let us begin by considering the closed string sector of the
background.  Melvin solutions supported by NSNS 3-form flux can be
constructed by applying the following sequence of dualities and twists.
\begin{enumerate}
\item Start with a flat background in type IIB supergravity
\be ds^2 = -dt^2 + dr^2 + r^2 d \phi^2 + dz^2 + \sum_{i=1,6} dy_i^2 \ee
where $z$ is compactified on a circle with radius $R$.
\item T-dualize along $z$ to obtain a background of type IIA supergravity
\be ds^2 = -dt^2 + dr^2 + r^2 d \phi^2 + d \tilde z^2 + \sum_{i=1,6} dy_i^2 \ee
where the radius of $\tilde z$ coordinate is $\tilde R = \alpha' / R$.
\item This geometry admits an isometry generated by a vector
${\partial / \partial \phi}$. Given such an isometry vector, one
can ``twist'' the compactification. By this, one means changing the
Killing vector associated with the compactification from $({\partial
/ \partial \tilde z})$ to $({\partial / \partial \tilde z} + \eta {
\partial / \partial \phi})$.  Alternatively, one can think of the
twist as first replacing
\be d\phi  \rightarrow d \phi + \eta \, d \tilde z \ee
so that the metric reads
\be ds^2 = -dt^2 + dr^2 + r^2 (d \phi + \eta d\tilde z)^2 + d \tilde z^2 + \sum_{i=1,6} dy_i^2 \label{IIAD2}\ee
and then treating $\tilde z$ as the periodic variable of radius $\tilde R$ with $\phi$ fixed.
\item T-dualize along $\tilde z$ to obtain a solution of type IIB supergravity
\beq
ds^2  & = &  -dt^2 + dr^2 + {r^2 \over 1 + \eta^2 r^2} d\phi^2 +  {1  \over 1 + \eta^2 r^2} d  z^2 + \sum_{i = 1}^6 dy_i^2  \cr
B & = & {\eta r^2 \over 1+\eta^2 r^2} d \phi \wedge d  z  \label{melvin} \\
e^\phi & = & \sqrt{{1 \over 1 + \eta^2 r^2}} \ . \nonumber
\eeq
\end{enumerate}
This solution describes a Melvin universe supported by a background NSNS
3-form flux. Our objective now is to consider the world volume theory
of D-branes in this background.  There is some freedom in the choice of
embedding of D-branes in this background.  In fact, precisely this
sequence of duality transformation in the presence of a D-brane was
used in the construction of dipole theories in
\cite{Bergman:2000cw,Bergman:2001rw}. In the context of dipole
theories, the D3-brane was embedded in such a way that it is extended
along the $(t, z, y_1, y_2)$ directions, but localized in the
$(r,\phi)$ plane. In other words, the isometry along which we twisted
the background corresponds to the R-symmetry from the point of view of
the D-brane world volume theory. In the appropriate scaling limit, the
theory on the world volume becomes a theory of dipoles whose length is
proportional to the R-charge.

Alternatively, one can embed a D3-brane in such a way that it is
extended along the $(t, r, \phi, z)$ and localized in the $y_i$
directions.  This time, the isometry with which we twist the
compactification is associated to the rotation along the world volume
of the gauge theory. One therefore expects to find a non-local theory
whose degree of non-locality is proportional to the angular momentum
quantum number. 

In order to read off the parameters appropriate for interpreting the
dynamics from the open string point of view, we apply the mapping of
Seiberg and Witten \cite{Seiberg:1999vs}
\be (G + {\theta \over 2 \pi \alpha'})^{\mu \nu} = [(g + B)_{\mu \nu}]^{-1} \ .  \label{swmap} \ee
When this formula is applied to the closed string background
(\ref{melvin}), one finds
\beq G_{\mu \nu} dx^\mu dx^\nu &=& -dt^2 + dr^2 + r^2 d \phi^2 + d  z^2 \cr
\theta^{\phi  z} & = & 2 \pi \alpha' \eta \label{openmetric}
\eeq
Therefore, in order to extract a field theory limit keeping the effect
of non-locality finite, one should scale $\eta$ so that
\be 2 \pi \alpha' \eta = 2 \pi \Delta = \mbox{finite} \ee
while sending $\alpha' \rightarrow 0$. We also keep the radius $R$ of
the periodic $z$ coordinate finite in this limit. In terms of the
Cartesian coordinates we have
\be \theta^{x_1 z} = -\theta^{z x_1}  = -2 \pi \Delta x_2, \qquad \theta^{x_2 z} =-\theta^{z x_2} = 2 \pi \Delta x_1 \label{melvintheta} \ee
with all other components vanishing. This is an example of
non-commutative gauge theory with non-constant non-commutativity for
which one needs to employ the formula of Kontsevich
\cite{Kontsevich:1997vb} to define the appropriate $*$-product. One
can readily verify that the condition
\be \theta^{il} \partial_l \theta^{jk}
+ \theta^{jl} \partial_l \theta^{ki}
+ \theta^{kl} \partial_l \theta^{ij} = 0 \ee
which is necessary for associativity, is satisfied by (\ref{melvintheta}). 

The formula (\ref{swmap}) of Seiberg and Witten was originally derived
for the case of a constant $B$-field background. One can nonetheless
see that the open string metric properly captures the effective
dynamics of open strings by observing that the induced metric
(\ref{IIAD2}) on the D2-brane after T-dualizing along the $z$
direction is identical to the open string metric found in
(\ref{openmetric}). The T-dual system consists of an
array of type IIA D2-branes at fixed positions in the periodic $\tilde z$
coordinate in the background space-time (\ref{IIAD2}).  Open strings
ending on these D2-branes will roughly follow geodesics along the
$t$-$r$-$\phi$ plane, which we see is identical to the geodesic of the
open string metric (\ref{openmetric}).  Further, T-duality along the $z$
direction does not affect the general features of the motion of these
strings along the $t$-$r$-$\phi$ coordinates.

One can also reach similar conclusions by observing that the Polyakov action
\be S = {1 \over 4 \pi \alpha'}\int d \tau \, d \sigma \left( \sqrt{-\gamma} \gamma^{\alpha \beta} \partial_\alpha X^\mu \partial_\beta X^\nu g_{\mu \nu} - \epsilon^{\alpha \beta} \partial_\alpha X^\mu \partial_\beta X^\nu B_{\mu \nu} \right) \ee
for open strings parameterized by strip coordinates $-\infty < \tau <
\infty$ and $0 < \sigma < \pi$, in the background (\ref{melvin}),
admits a  solution of the equation of motion, constraints, and the boundary condition of the form
\beq t(\tau,\sigma) & = & \left({L \over \pi \eta b}\right) \tau \cr
\rho(\tau,\sigma) & = & \sqrt{b^2 + t(\tau,\sigma)^2} \cr
\phi(\tau,\sigma) & = & \tan^{-1} \left({t(\tau,\sigma) \over b}\right) \cr
z(\tau,\sigma)  & = & {L \over \pi} \sigma \cr
y_i(\tau,\sigma)  & = & y_i \cr
\gamma_{\alpha \beta}(\tau,\sigma) & = & \eta_{\alpha \beta} \ . 
\eeq
This solution describes the motion of an open string which is extended like a rod along the
$z$ direction.  The string is traveling on a straight line in the $r$-$\phi$ plane with impact parameter $b$ and is propagating at the speed of light with respect to the open string metric.  This is further evidence supporting the relevance of $G_{\mu \nu}$ for the effective dynamics of open strings. The angular
momentum of this configuration is
\be J = \int d \sigma {\partial {\cal L} \over \partial  (\partial_\tau \phi)} = {L \over 2 \pi \alpha' \eta} \ . \ee
Combining this result with (\ref{openmetric}), one can easily verify
that the length of the rod is is proportional to  angular momentum 
\be L = \theta^{\phi z} J \ee
with the non-commutativity parameter being the constant of
proportionality.  The fact that the open strings become dipoles of
length $L$ gives rise to non-locality in their interaction, precisely
of the type that one would expect for the non-commutativity between
angular coordinate $\phi$ and a Cartesian coordinate $z$.

String theory in space-times generated by acting with Melvin twists
are implicitly simple. It is possible to quantize the strings and to
construct explicit boundary states describing D-branes in these
backgrounds \cite{Dudas:2001ux,Takayanagi:2001aj}.  One can also take
advantage of intrinsic simplicity to derive the supergravity dual of
the non-commutative gauge theory which arise as a decoupling limit of
D3-brane in (\ref{melvin}). Simply start with the supergravity
solution of the D3-brane
\beq ds^2 &=& f(\rho)^{-1/2}(-dt^2 + dr^2 + r^2 d \phi^2 + dz^2) + f(\rho)^{1/2}(d\rho^2 + \rho^2 d \Omega_5^2), \cr
f(\rho) &=& 1 + {4 \pi g N \alpha'^2  \over \rho^4} \eeq
and follow the chain of dualities outlined in this section. Finally, scale
\be \rho = \alpha' U, \qquad \eta = {\Delta \over \alpha'} \ee
and send $\alpha' \rightarrow 0$ keeping $U$ and $\Delta$ fixed. The resulting geometry is given by a metric which in string frame has the form
\be ds^2 = \alpha' \left( {U^2 \over \sqrt{\lambda}} \left(-dt^2 + dr^2 + {r^2 d \phi^2 + dz^2 \over 1 +  {\Delta^2 r^2 U^2 \over \lambda}} \right) + {\sqrt{\lambda} \over U^2}(dU^2 + U^2 d \Omega_5^2) \right) \ . \label{sgdual} \ee
This is essentially the construction used in
\cite{Hashimoto:2002nr,Bergman:2001rw} with minor difference in the
orientation of the brane and the isometry of the twist. The resulting
space-time is the supergravity dual of the non-commutative gauge
theory with non-commutativity (\ref{melvintheta}) along the lines of
\cite{Hashimoto:1999ut,Maldacena:1999mh}.

\section{Other Melvin twists and non-commutative gauge theories}

In the previous section, we described the construction of
non-commutative gauge theory with the non-constant non-commutativity
indicated in (\ref{melvintheta}). Essentially, this is a simpler
version of the construction of a theory with time dependent
non-commutativity parameter outlined in \cite{Hashimoto:2002nr}.  In
this section, we will examine if other non-commutative gauge theories
admit a realization as the world volume dynamics of D-branes in a closed
string background which is a Melvin twist of flat space.

One non-commutative gauge theory which has been studied by Dolan and
Nappi \cite{Dolan:2002px} is the world volume theory of D-branes in the
Nappi-Witten background \cite{Nappi:1993ie}.  The Nappi-Witten background
can be viewed as a 3+1 dimensional plane wave supported by a null
background NSNS 3-form flux. It can be generated by performing the
Melvin twist along the light-like direction, or equivalently, by
combining the chain of dualities leading to the construction of the Melvin
universe (\ref{melvin}) with a boost.  Following \cite{Gimon:2003xk}
we refer to this sequence of solution generating transformations as the
{\it null Melvin twist}. Detailed explanation of steps and scalings
involved with the null Melvin twist can be found in
\cite{Alishahiha:2003ru,Gimon:2003xk}. The Null Melvin twist applied to 
flat the background of type IIB supergravity gives rise to a solution
\beq ds^2 & = & -dt^2 + d z^2 - 2 \beta^2 r^2 (dt + d z)^2 + dr^2 + r^2 d \phi^2 +  \sum_{i = 1}^6 dy_i^2 \cr
B & = &  \beta r^2 d \phi \wedge (dt +  dz) \label{planewave} \\
e^\phi & = & 1 \nonumber \ . \eeq
Now consider a D3-brane extended along $(t,r, \phi, z)$ and localized
along the $y_i$ directions.  It is not difficult 
apply the  Seiberg-Witten map (\ref{swmap}) to show that the open string metric and the non-commutativity parameter are given by
\be G_{\mu \nu} dx^\mu dx^\nu = -dt^2 + dr^2 + r^2 d \phi^2 + d  z^2 , \qquad
 -\theta^{\phi t} = +\theta^{t \phi}  =   \theta^{\phi z} = - \theta^{z \phi} = 2 \pi \alpha' \beta \ .  \label{nullmelvin}
\ee
One can therefore obtain a non-commutative gauge theory with finite non-commutativity parameter by taking the limit $\alpha' \rightarrow 0$ keeping
\be \Delta = \alpha' \beta = \mbox{fixed} \ . \ee 
In Cartesian coordinates, we have
\be -\theta^{x_1 t} = \theta^{t x_1} = \theta^{x_1 z} = -\theta^{z x_1}  = -2 \pi \Delta x_2, \qquad - \theta^{x_2 t} = \theta^{t x_2} = \theta^{x_2 z} =-\theta^{z x_2} = 2 \pi \Delta x_1 \label{nullmelvintheta}\ . \ee
This non-commutativity is non-constant, but not time dependent. This
is different from $t$ dependent non-commutativity which was reported
in \cite{Dolan:2002px}. This difference can be attributed to the
difference in the coordinates and the gauge for the NSNS 2-form of the
closed string background. To see this more explicitly, note that the
Nappi-Witten background used in \cite{Dolan:2002px} used the
coordinate
\be \phi = \phi' + \beta (t + z) \label{coorchange} \ee
so that the metric takes the form
\be ds^2  =  -dt^2 + d z^2 - 2 \beta r^2 (dt + dz) d \phi' + dr^2 + r^2 d \phi'^2 +  \sum_{i = 1}^6 dy_i^2 \ee
while the $B$ field  differs from (\ref{planewave}) by a total
derivative term
\be B' = \beta r^2 d \phi' \wedge (dt + dz) + d (\beta r^2(t+z) d \phi') = 2 \beta  r (t+z) \,  dr \wedge d \phi' \ . \label{bprime}\ee
In this gauge, the $B'$ field has explicit dependence on $t$ which is
suggestive of a time dependent non-commutativity.

However the equation of motion for the gauge field derived from
the DBI action for a general closed string background
\cite{Leigh:1989jq} does not admit $F=0$ as a solution for
(\ref{bprime}). So (\ref{bprime}) with trivial gauge field $F=0$ is
not a consistent background of string theory.  On the other hand,
(\ref{planewave}) does admit $F=0$ as a consistent solution to the
equation of motion \cite{Stanciu:1998ak}. We are therefore led to
conclude that the open string metric and the non-commutativity
parameter in the decoupling limit of D3-branes in Nappi-Witten
background are that of (\ref{nullmelvin}).

It is interesting to note that other known examples of non-commutative
gauge theory can be viewed as being generated by a sequence of
Melvin-like twists.  For example, the standard non-commutative gauge theory with constant space-like non-commutativity can be thought of as being generated via a sequence where one
\begin{enumerate}
\item Start with flat space
\be ds^2 = -dt^2 + dx^2 + dy^2 + dz^2 + d \rho^2 + \rho^2 d \Omega_5^2 \ee
\item T-dualize the compact coordinate $z$ with radius $R$ so that it
becomes a compact coordinate $\tilde z$ with radius $\alpha' / R$.
\item Twist by replacing $dy \rightarrow dy + ({\Delta^2/ \alpha'}) d \tilde z$
\item T-dualize back on $\tilde z$.
\end{enumerate}
This will give rise to a constant NSNS B-field which via the
Seiberg-Witten map gives rise to a theory with non-commutativity
$\theta^{yz} = 2 \pi \Delta^2$. Because this amounts to twisting along
the isometry $\partial/ \partial y$ which corresponds to shift in
the $y$ direction, one might refer this sequence of dualities as {\it
Melvin shift twist}.

Combining the Melvin shift twist with a boost along the $z$ coordinate
will give rise to a non-commutative gauge theory with light-like
non-commutativity \cite{Aharony:2000gz}. Since this is equivalent to
T-dualizing along the null direction accompanied by a twist by shift
along the $y$ coordinate, one might refer to this construction as the
{\it null Melvin shift twist}.

One might also consider combining T-duality along the $z$ direction
and a twist with respect to boost along the $x$ coordinate.  This will
give rise to a background which is T-dual to the background considered
in \cite{Cornalba:2002fi}. However, the fact that $\tilde z$
coordinate is time-like in some region gives rise to pathology which
makes interpretation of the decoupled theory as a non-commutative
theory unreliable. This can be cured by twisting by a combination of a
boost and a rotation so that the net effect is twisting by a null
rotation.  An appropriate name for this sequence of dualities and
twists is the {\it Melvin null twist}. This gives rise to a closed
string background which is a T-dual of the null brane
\cite{Figueroa-O'Farrill:2001nx,Liu:2002kb}. In the presence of
D3-brane, this gives the construction of \cite{Hashimoto:2002nr}. (The
S-dual corresponding to NCOS with time dependent non-commutativity was
considered in \cite{Cai:2002sv}.)

The non-commutative gauge theories generated along these lines are
summarized in table \ref{table1}.  It is interesting to observe that
most of the known examples of non-commutative gauge theories can be
thought of as being generated by sequences of dualities and twists. We
have also included twists along the direction transverse to the brane
which we refer to as the R-twist.

\begin{table}
\centerline{\begin{tabular}{|l|l|} \hline
Type of Twist & Model \\ \hline \hline 
Melvin Twist & Model of section 2 \\ \hline
Null Melvin Twist & Dolan-Nappi model \\ \hline
Melvin Shift Twist & Moyal model \\ \hline
Null Melvin Shift Twist & Aharony-Gomis-Mehen model\\\hline
Melvin Null Twist & Hashimoto-Sethi model\\ \hline
Melvin R Twist & Bergman-Ganor model \\ \hline
Null Melvin R Twist & Ganor-Varadarajan model \\ \hline
\end{tabular}}
\caption{Catalog of non-commutative gauge theories viewed as a world
volume theory of D-branes in a ``X'' Melvin ``Y'' twist
background. \label{table1}}
\end{table}

A notable omission in this catalog is the model of
Lowe-Nastase-Ramgoolam \cite{Lowe:2003qy}. From the point of view of
constructing a non-commutative gauge theory, it is convenient to think
of this setup as placing a D3-brane probe in the background of smeared
NS5-branes. While this is a perfectly sensible closed string
background that one can consider, it appears not to be related to flat
space under any chain of dualities.

\section{World volume theory of branes in a decaying Melvin background}

Each of the non-commutative gauge theories enumerated in Table 1 are
interesting in their own right. The example of embedding D3-branes in a 
Melvin universe, however, is special in that the closed string
background breaks supersymmetry and is susceptible to decay.  There
are several decay modes which arise as double wick rotation of black
holes solutions, originally considered in
\cite{Dowker:1994bt,Dowker:1994up,Dowker:1995gb,Dowker:1996sg} and
more recently in the context of string theory in \cite{Costa:2000nw}.
  It would be interesting to see what
kind of effective open string dynamics arise as a decoupling limit of
D-branes placed in this class of time dependent background.  It is
natural to expect that the world volume theory will inherit the time
dependence of the background space-time, giving rise to a new type of
non-commutative gauge theory with explicit time
dependence.

Since we are interested in the interplay between background NSNS
3-form flux and the non-commutativity on the world volume of
D3-branes, it is natural to consider the decay mode via nucleation of
monopole anti-monopole pair which subsequently fly off to infinity.
To be more concrete, we consider a background where a monopole and an
anti-monopole, which are smeared along $y_2 \ldots y_6$ coordinates,
undergoes uniform acceleration along the $y_1$ direction.  This is
precisely the process which was considered in detail in
\cite{Emparan:2001gm}.  Our interest is in exploring the effect of
the accelerating monopole anti-monopole pair on the world volume theory of
D3-brane extended along $(t,r,\phi,z)$ coordinates and localized half
way between the monopole anti-monopole at $y_1 = 0$.

Let us review the decaying Melvin solution more explicitly. The most efficient way to describe this background is to start from the Euclidean Kerr instanton solution in 5 dimensions
\beq ds^2 &=& r^2 \cos^2 \theta\,  d\psi^2 + (r^2 - \alpha^2 \cos \theta^2)\,  d \theta^2 + {r^2 - \alpha^2 \cos \theta^2 \over r^2 - \alpha^2 - \mu}\,  dr^2 \cr
&& + dz^2 + \sin^2 \theta (r^2-\alpha^2) d \phi^2 - {\mu \over r^2 - \alpha^2 \cos^2 \theta} (dz + \alpha \sin^2 \theta d \phi)^2 \eeq
and do a Wick rotation $\psi = it$
\beq ds^2 &=& -r^2 \cos^2 \theta\,  dt^2 + (r^2 - \alpha^2 \cos \theta^2)\,  d \theta^2 + {r^2 - \alpha^2 \cos \theta^2 \over r^2 - \alpha^2 - \mu}\,  dr^2 \cr
&& + dz^2 + \sin^2 \theta (r^2-\alpha^2) d \phi^2 - {\mu \over r^2 - \alpha^2 \cos^2 \theta} (dz + \alpha \sin^2 \theta d \phi)^2 \ . \label{KerrErnst} \eeq
This geometry covers the $(t,r,\phi,z,y_1)$ plane in different coordinates.
We would like to consider the $z$ coordinate to be compact with period
$2 \pi R$.  The Euclidean horizon velocity of this metric is
\be \Omega = - {g_{\phi z} \over g_{\phi \phi}} = {\alpha \over \mu} \ . \ee
It is therefore natural to compactify along the Killing vector $(d/dz) + \Omega (d/d \phi)$. One can equivalently twist the coordinates
\be z = \tilde z, \qquad \phi = \tilde \phi + \Omega \tilde z \ee
and compactify along $(d/d\tilde z)$. Generically, there will be a conical singularity in the $r$-$\tilde z$ plane, unless the period of the compactification is adjusted according to the parameter of the Kerr instanton solution. This is achieved by choosing the radius of compactification according to
\be \tilde z = \tilde z + R, \qquad R^2 = {\mu^2 \over \mu + \alpha^2} \ . \ee
This is the analogue of the ``bubble of nothing'' for the Melvin
universes \cite{Witten:1981gj}. One can actually consider more general
twists
\be z = \tilde z, \qquad \phi = \tilde \phi + \left(\Omega + {n \over R}\right) \tilde z \ . \ee
The case $n=1$ is what is referred to as the ``Shifted Kerr''
solution. This is the solution which corresponds to the Ernst solution in
different coordinates as was shown in \cite{Dowker:1995gb}.  Since the
original Kerr instanton is asymptotically flat, its Lorentzian
continuation is also flat at large spatial distances away from the
origin.  The dimensional reduction of this 4+1 dimensional geometry
along a twisted $z$ coordinate will therefore at large distances give rise to a Melvin universe supported by the flux of the Kaluza-Klein gauge field.  The presence of the Kerr instanton will manifest
itself as a Kaluza-Klein monopole anti-monopole pair flying away in
this background Melvin geometry.  The twist parameter $\eta$ for this
geometry at spatial infinity is
\be \eta = {\alpha \over \mu} + {1 \over R} \ . \ee

This 4+1 dimensional geometry can be embedded into type IIA or type
IIB supergravity in 10 dimension by simply adjoining 5 flat
directions. We will consider the case of IIA where the Kaluza-Klein monopole is
now a five-brane. T-dualizing this geometry along $\tilde z$ will give
rise to an asymptotically Melvin solution supported by NSNS $B$-field in
IIB supergravity.  This is the same Melvin geometry considered in
section 2. The IIA Kaluza-Klein monopole becomes NSNS 5-brane extended along the $y_2$-$y_6$ directions and smeared
along the $\tilde z$ direction in this picture.

Roughly speaking, the monopole anti-monopole pair partially screens
the Melvin flux.  The natural place to probe this effect is along the
plane which is half way between the monopole and the
anti-monopole. Unfortunately, the coordinates used in writing
(\ref{KerrErnst}) is not adequate for this purpose for it does not
cover the entire geometry and misses this important region. The fact
that the coordinates of (\ref{KerrErnst}) is incomplete can be seen by
noting that for large $r$, $t$-$\theta$-$\phi$ plane describes de
Sitter space in static coordinates, which is geodesically
incomplete. This problem is easy to rectify.  Simply go to the global de Sitter
coordinates by mapping
\be 
t  =  \sinh^{-1} \left( \sinh \tilde t \over \sqrt{1 - \cosh^2 \tilde t \sin^2 \tilde \theta} \right) , \qquad 
\theta = \sin^{-1} ( \cosh \tilde t \sin \tilde \theta )  \ . 
\ee
Now, the plane bisecting the monopole anti-monopole pair can be
identified as the surface $\tilde \theta=\pi/2$.

\begin{figure}
\centerline{\includegraphics[width=2.5in]{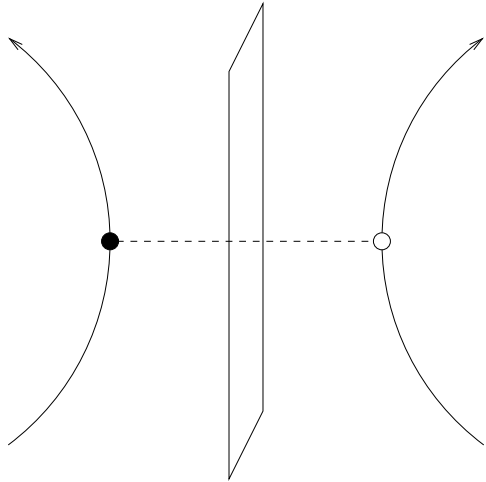}}
\caption{Configuration of D3-brane oriented along the plane bisecting the monopole anti-monopole pair in a decaying Melvin universe}
\end{figure}

At this point one can T-dualize along $\tilde z$, place the D3-brane
at $\tilde \theta = \pi/2$, and study the effective open string
dynamics.  It can be checked that the ansatz $\tilde \theta=0$,
$y_i=0$ and vanishing world volume gauge field is consistent with the
equation of motion of the DBI action in this background.  In order to take the decoupling limit, we will scale
\be \eta = {\Delta \over \alpha'}, \qquad R = {2 \pi \alpha' \over L} \label{scaling} \ee
for fixed $\Delta$ and $L$ as we send $\alpha'$ to zero. The scaling
of $R$ is necessary in order for the $z$ coordinate in the IIB picture
to have finite period $z \sim z + L$. This will cause $\mu$ and
$\alpha^2$ to scale like $\alpha'^2$.  

Our task now is to extract the open string metric and the
non-commutativity parameter in this scaling limit. As we found before,
naive application of Seiberg-Witten formula (\ref{swmap}) gives the
same metric in the $\tilde t$-$r$-$\phi$ plane as the IIA metric in the
T-dual picture, ensuring that the open string metric properly captures
the geodesic motion  of the open strings.

Along the $\tilde t$-$r$ coordinates, the induced metric on the D3-brane takes the form
\be (r^2 + \alpha^2 \sinh^2 \tilde t) \left(-d \tilde t^2 + {dr^2 \over r^2 - \mu - \alpha^2} \right) \ . \ee
At large $r$, this looks like a Rindler coordinate.  However, this coordinate ends at $r^2 = \mu + \alpha^2$ and the Rindler horizon is not encountered. Instead, the region near $r^2 = \mu + \alpha^2$ is actually regular.  To exhibit this feature, it is useful to go to the null coordinates
\be u = t + r_*, \qquad v = t - r_* \ee
where
\be r_* = \int dr \sqrt{{g_{rr} \over g_{tt}}} = \log\left({r + \sqrt{r^2 - \mu - \alpha^2} \over \sqrt{\mu + \alpha^2}}\right) \ , \ee
and make the standard transformation to take the Rindler coordinate into Cartesian coordinates
\be u = \log\left({U \over \sqrt{\mu + \alpha^2}}\right), \qquad v = -\log\left(-{V\over \sqrt{\mu + \alpha^2}}\right) \ . \ee
The cut-off $r^2 = \mu+\alpha^2$ now corresponds to
\be UV = -(\mu+\alpha^2) \ee
which is a time-like curve. This time-like curve can be mapped to curve $x=0$ parameterized by $\tau$ by making further coordinate transformations
\be U = \tilde U + \sqrt{\tilde U^2 + \mu + \alpha^2}, \qquad V = \tilde V - \sqrt{\tilde V^2 + \mu + \alpha^2} \ee
and 
\be \tilde U = \tau + x, \qquad \tilde V = \tau-x \ . \ee
In these coordinates, the closed string metric in the neighborhood of $x=0$ looks like
\be ds^2 =  \left( \frac{{\alpha}^4 + {\mu}^2 +
       {\alpha}^2\,\left( 2\,\mu + {\tau}^2 \right) }{\left( {\alpha}^2 +
         \mu \right) \,\left( {\alpha}^2 + \mu + {\tau}^2 \right) } \right) \left(-d \tau^2  + dx^2  + x^2 d \phi^2 +dz^2\right) \ee
which is perfectly regular.

Now we are ready to compute the open string parameters. The scaling (\ref{scaling}) dictates that 
\be q^2 \equiv {\alpha^2 \over \mu} = {(L - 2 \pi \Delta)^2 \over 4 \pi \Delta (L - \pi \Delta)} \ee
is fixed and in the decoupling limit, the open string metric develops a discontinuity: we find
\be ds_{open}^2 = \left(1 - {\Theta(\tau^2-x^2) \over 1 + q^2} \right)(-d\tau^2 + dx^2) + x^2 d \phi^2 + d z^2 \ee
where $\Theta(x)$ is the step function.  The non-commutativity parameter comes out to 
\be \theta^{\phi z} = 2 \pi \Delta \ . \ee
This appears to describe a non-commutative gauge theory with additional explicit time dependence in the form of the metric.   The geometry is almost flat. For $x^2 - \tau^2 > 0$, it is identical to the non-commutative gauge theory of section 2. Perhaps this is to be expected since in the decoupling limit, both $\mu$ and $\alpha$ is going to zero.

\section{Conclusions}

In this article, we examined the world volume theory of D3-branes
wrapping a Melvin universe supported by NSNS $B$-field and found that
it describes a non-commutative gauge theory with non-constant
non-commutativity in the appropriate scaling limit.  Melvin universes
are particularly simple in that they can be gotten from applying
dualities and twists on flat space.  The particular sequence of
transformations leading to the Melvin universe is called the Melvin
twist.  We find that many examples of non-commutative guage theories
with non-constant non-commutativities can be generated from a slight
variation of the Melvin twist. The model of Hashimoto and Sethi for
example arises from Melvin null twist, whereas the model of Dolan and
Nappi can be realized as the null Melvin twist.  Melvin twists appear
to provide a unified perspective on most of the explicitly known
construction of non-commutative gauge theory as a decoupled theory on
branes.

We also studied the world volume theory of D-brane embedded into
Melvin universe decaying via nucleation of monopole anti-monopole
pair. Such a background exhibits rich dynamics in the closed string
sector and was investigated in detail in
\cite{Emparan:2001gm}. Unfortunately, most of the time dependence
appear to be smoothed out in the process of taking the decoupling
limit of the world volume gauge theory. The only remnant of the time
dependence we find is a discontinuity in the open string metric along
the light-cone.  Perhaps the physics of open strings where one does
not completely decouple the closed strings and excited states will
exhibit more interesting dynamics.

Models constructed using Melvin twists are particularly simple.  The
world sheet theory for strings in these background are exactly
solvable and so it should be possible to make many precise statements
about these theories from the string theory point of view. It is also
possible to write down an explicit action for these non-commutative
gauge theories and analyze perturbative issues as was done in
\cite{Cerchiai:2003yu,Robbins:2003ry}. It would also be interesting to
explore the instanton, monopole, and vortex solutions of
non-commutative gauge theories
\cite{Nekrasov:1998ss,Hashimoto:1999zw,Gross:2000wc,Gross:2000ph,Aganagic:2000mh,Harvey:2000jb,Gross:2000ss,Hashimoto:2000ys},
and to study the structure of gauge invariant operators
\cite{Das:2000md,Gross:2000ba}, along the lines of what was done for
the theories with constant non-commutativity parameters.

\section*{Acknowledgments}

We would like to thank
J.~Gauntlett,
N.~Itzhaki,
H.~Nastase, and
T.~Takayanagi
for useful discussions, and
L.~Dolan and C.~Nappi for correpondence.  We would also like to thank
the IAS and especially the organizers of PiTP 04 during which part of
this work was done. This work was supported in part by the funds from
the University of Wisconsin.

\bibliography{twist}\bibliographystyle{utphys}

\end{document}